\documentclass[a4paper,11pt]{article}%
\usepackage[utf8]{inputenc}
\usepackage{amsmath}
\usepackage{physics}
\usepackage{bbold}
\usepackage{graphicx}%
\usepackage{comment}
\usepackage{multicol}
\usepackage{booktabs}
\usepackage{xcolor}
\usepackage{listings}
\usepackage{upquote}
\input{lst-color.sty}
\usepackage{microtype}
\usepackage{wasysym}
\usepackage{units}
\usepackage{fancyvrb} 
\usepackage{xspace}   
\usepackage{slashed}
\usepackage{environ}
\usepackage{framed}

\usepackage{pxfonts}

\usepackage[autostyle]{csquotes}
\usepackage{cite}

\usepackage{url} 
\usepackage{hyperref} 
\usepackage{xurl} 

\newcommand{\ped}[1]{_{\scriptscriptstyle{\textrm{#1}}}}
\newcommand{\peff}{p\ped{eff}}

\newcommand{\gstrong}{g\ped{s}}
\newcommand{\alphastrong}{\alpha\ped{s}}
\newcommand{\Weff}{W\ped{eff}}
\newcommand{\TeV}{{\rm TeV}}
\newcommand{\MadDM}{\protect\Verb|MadDM|\xspace}
\newcommand{\DarkSUSY}{\protect\Verb|DarkSUSY|\xspace}
\newcommand{\MicrOMEGAs}{\protect\Verb|MicrOMEGAs|\xspace}
\newcommand{\Madgraph}{\protect\Verb|Madgraph|\xspace}
\newcommand{\CalcHEP}{\protect\Verb|CalcHEP|\xspace}
\newcommand{\SuperIsoR}{\protect\Verb|SuperIso Relic|\xspace}
\newcommand{\SuperIsoRF}{\protect\Verb|SuperIso Relic v4.1|\xspace}
\newcommand{\SuperIso}{\protect\Verb|SuperIso|\xspace}
\newcommand{\MARTY}{\protect\Verb|MARTY|\xspace}
\newcommand{\FORTRAN}{\protect\Verb|FORTRAN|\xspace}
\newcommand{\DarkPack}{\protect\Verb|DarkPACK|\xspace}
\newcommand{\FormCalc}{\protect\Verb|FormCalc|\xspace}

\title{DarkPack: A modular software to compute BSM squared amplitudes for particle physics and dark matter observables}
\author{}
\date{}

\begin{document}

\hfill {\tt CERN-TH-2022-197}

\def\thefootnote{\fnsymbol{footnote}}

\begin{center}
{\Large
\vspace{0.3cm}
\DarkPack: a modular software to compute BSM squared amplitudes for particle physics and dark matter observables}

\vspace{0.3cm}
{\large\bf  
M.~Palmiotto$^{a}$\footnote{Email: marco.palmiotto@univ-lyon1.fr},
A.~Arbey$^{a,b}$,
F.~Mahmoudi$^{a,b}$
}
 
\vspace{0.3cm}
{\em $^a$Universit\'e de Lyon, Universit\'e Claude Bernard Lyon 1, CNRS/IN2P3, \\
Institut de Physique des 2 Infinis de Lyon, UMR 5822, \\F-69622, Villeurbanne, France}\\[0.2cm]
{\em $^b$Theoretical Physics Department, CERN, CH-1211 Geneva 23, Switzerland} 

\end{center}

\renewcommand{\thefootnote}{\alph{footnote}}
\setcounter{footnote}{0}

\thispagestyle{empty}

\begin{abstract}
    We present here a new package to automatically generate a complete library of 2 to 2 squared amplitudes at leading order in any New Physics model. The package is written in \texttt{C++} and based on the \MARTY software. 
    The numerical library generated allows for the computation of relic density by embedding the algorithms of \SuperIsoR.
\end{abstract}

\tableofcontents

\section{Introduction}
The question of the nature of dark matter is an important topic in both particle physics and cosmology (see \cite{Arbey:2021gdg} for a recent review). For decades, the most studied New Physics (NP) scenario providing a dark matter candidate has been the minimal supersymmetric extension of the Standard Model (MSSM), for which several softwares have been developed to compute dark matter observables. However, so far no supersymmetric particle has been discovered in particle physics experiments, and other NP scenarios are worth being investigated. In this context, the software \SuperIsoR 
\cite{Arbey:2009gu,Arbey:2011zz,Arbey:2018msw}, which aims at studying dark matter observables in the standard cosmological model and in alternative cosmological scenarios, has been until now focused on the MSSM and NMSSM, 
and is being developed in order to study other NP scenarios.
To achieve this purpose, it needs a new way of handling the computation of matrix elements, since it relies on a self-generated \FormCalc \cite{Hahn:2016ebn} code, therefore not meant to be particularly flexible.

An effort in the direction of generalising and automating the computations has already been performed through the development of the code \MARTY \cite{Uhlrich:2020ltd}, which allows for the automatic calculation of amplitudes, cross-sections and Wilson coefficients in a large variety of NP models up to the one-loop level. However, a step further is needed to compute dark matter observables, such as dark matter relic density, which can involve thousands of scattering amplitudes.

Other codes publicly available to calculate dark matter observables are \DarkSUSY \cite{Gondolo:2004sc,Bringmann:2018lay}, \MadDM \cite{Ambrogi:2018jqj,Arina:2020kko} and \MicrOMEGAs \cite{Belanger:2018ccd}. Despite its name \DarkSUSY is model-independent. In fact, provided some inputs such as the DM mass and its self-annihilation cross-section, it is possible to compute the relic density, alongside other observables related to direct or indirect detection of dark matter. Moreover, it has a modular structure that makes it easy to link with other tools. 

\MadDM and \MicrOMEGAs are also model-independent: the former relies on \Madgraph \cite{Alwall:2014hca}, and the latter on \CalcHEP \cite{BELYAEV20131729}. 
It is therefore possible to provide model inputs compatible with those softwares and then all the relevant quantities for relic density calculation, 
as for instance the annihilation cross sections, are computed from their respective backends, 
in order to compute DM relic density and other DM related observables.
In order to test a new model in any of these frameworks, the generation of the necessary input files would have to 
be performed with \lstinline{Sarah} \cite{Staub:2013tta}. 
Also, \MadDM is the only tool that can easily provide results accurate up to $n$ loops with custom models, while the other ones stop at the tree level.\footnote{
Some of them can have one- or multi-loop accuracy for the MSSM.}

In this paper, we present the package \DarkPack, which aims at automatically generating a numerical library of scattering amplitudes to compute dark matter observables such as relic density, and which is interfaced with \SuperIsoRF. 
With \DarkPack, we want to propose a tool that can handle by itself all the steps from the definition of the model to the output of the sum of the squared amplitudes and the computation of dark matter related observables. 
\DarkPack is intended to be comprehensive, in the sense that we avoid unnecessary passages of input and output.
Among the advantages of this approach, there is a larger flexibility in the code usage.
In fact, it is easier to modify an algorithm if it is written in a pre-established comprehensive framework.
For such reasons we have chosen to rely on \lstinline{MARTY} for the model building and 
the symbolic manipulations.
Nevertheless, it is crucial to have modularity as a feature, since this enables an easy linking with external 
softwares if needed.

In \DarkPack \lstinline{v1.1} the generated library contains all the squared amplitudes of 2 NP particles into 2 Standard Model (SM) particles at the leading order and up to the one-loop level.
In order to validate the results, we focus on the MSSM which is an adequate benchmark in terms of complexity and compare in particular the results of \DarkPack with those of \SuperIsoRF, which relies on a self-generated 
\FormCalc \cite{Hahn:2016ebn} \FORTRAN code.\footnote{
As explained in the conclusion, we foresee an upgrade of \SuperIsoR based on \DarkPack, since on one side
\DarkPack is conceived to be modular, and on the other side the way \SuperIsoRF computes amplitudes is 
not suited to be flexible and adaptable to compute new observables.
}

This article is structured as follows. In section \ref{sec:setup} the compilation instructions of the package are provided.
In section \ref{sec:outlook} we present the philosophy behind the code, i.e.~the main goals and the methods used. Furthermore, we describe the role of \MARTY and of the numerical library, as well as general information such as how to 
handle numerical inputs and where to find the example programs we provide.
The main functions of \MARTY that are used to generate the numerical library are given in section \ref{sec:marty}.
In section \ref{sec:numerical} we describe how to use the numerical library. Functions are listed accordingly to the user's needs. 
Further information can be found in the appendices and in the source code, where the functions are described in the comments.

\section{Setup of the package}\label{sec:setup}

\DarkPack \lstinline{v1.1} is available among the attachments of this article.
Alternatively, the source code is hosted on the GitLab repository of the IN2P3, and 
it can be found at the address:\\
\url{https://gitlab.in2p3.fr/darkpack/darkpack-public}.\\
The version \lstinline{1.1}, described in this work is available as a commit 
in the \lstinline{master} branch with tag \lstinline{v1.1}.

In order to use the code, the user needs a working setup of \MARTY installed\footnote{We provide a script that automatically installs \MARTY. More information about the installation and usage of \MARTY can be found at \url{https://marty.in2p3.fr} and in Refs.~\cite{Uhlrich:2020ltd,Uhlrich:2020aaj,Uhlrich:2021fwl,Uhlrich:2021ded}.}. No further dependencies are needed. 
To compile the library, the user has to run the script \texttt{lib\_setup.sh}
providing two mandatory arguments and two optional arguments:
\begin{itemize}
    \item The name of the numerical library\footnote{by default \texttt{mssm2to2} for the MSSM.}, that we denote below as \lstinline{<1>}.
    \item The number of threads used for the compilation (use 1 in case of doubt), or the string \lstinline{-nomake} for not compiling automatically. 
    Specifying the number of threads, the type of the default linking is determined by the value variable \lstinline{linking}, defined at the beginning of the script. 
    \item Optionally, the name of another numerical library as a fourth argument, preceded by a \lstinline{-R} as the third argument. This has to be used when compiling a new library for the first time, while the auxiliary files have been used to compile another library previously. By specifying this argument, a search and replace is performed on the files to update them with the new library name.\footnote{
    We point out that the search and replace happens for any occurrence of the string given as input here.
    Hence, we strongly advise to use numerical library names which are very unlikely to appear in other parts of the code. As a general rule of thumb, adding a suffix, such as \lstinline{2to2} for the MSSM, can be an effective solution.}
\end{itemize}
The script will work by using the files in \lstinline{auxiliary_library/<1>} as the source code for the extra functions specific to the library and the files in \lstinline{auxiliary_library/script_<1>} as the source code for the executable files. For standard use cases, the files in these two directories are the ones to be written specifically for any new library. Please see the content of section \ref{sec:numerical} for the description of the source files we provide, both general and library-specific. 

The content of the package is the following. The numerical library for the MSSM can be found in the folder \texttt{mssm2to2}, and external files for the library in the \lstinline{auxiliary_library} folder. 

In order to generate the library from scratch\footnote{As just pointed out, the MSSM library is provided within the package and is not required to be generated.},  the source file (e.g.~\texttt{MSSM.cpp} for the MSSM) has to be compiled and the corresponding program launched.

For this purpose, the user can run the script \lstinline{./lib_generate.sh}, providing as arguments the name of the target library and the name of the source code without the \lstinline{.cpp} extension:
\begin{verbatim}
 ./lib_generate.sh <library name> <source code name>
\end{verbatim}
and then the generated library has to be compiled with the instructions of the previous paragraph.

Further instructions can be found in the \texttt{README.md} file.

\section{Outlook of the code}\label{sec:outlook}

This code has been developed with the purpose of having a new tool to compute 2 to 2 scattering amplitudes, firstly in the MSSM at the leading order (LO). 
In this release, we provide the setup for the MSSM.\footnote{In a future release we will provide the setup for a model with a scalar DM candidate and a scalar messenger (see e.g.~ \cite{Buckley:2014fba,Abercrombie:2015wmb}) as an example.}

The idea behind this package is to use \MARTY in order to define the desired model and quantities to compute and to export a numerical \texttt{C++} library. 
With \MARTY it is possible to compute symbolically scattering amplitudes up to the one-loop level.
However, \texttt{MARTY 1.6} does not provide tools to apply automatically a renormalization scheme.
As a consequence, a computation can be successfully performed at one loop if 
there are no divergent diagrams at the tree level, meaning that the one-loop level has to be the leading order.\footnote{
In the process list that we provide there are no processes at 1 loop because their symbolic computation 
in models with a large number of fields, such as the MSSM, takes a lot of time.
Furthermore, as will be explained later, we used the tree level in order to be able to compare the numerical results 
with \SuperIsoRF,
which relies on a self-generated \FormCalc code.}
In particular, in this release there is the example file
\texttt{MSSM.cpp} which contains the code that computes symbolically all the sums of 2 to 2 squared amplitudes
relevant for the computation of relic density in the MSSM. 
The process list we consider is the same as in \SuperIsoRF and can be found in the file \lstinline{data/processes_all.psiso}. 
The list can be easily extended in many ways. 
For instance, it is possible to add manually some processes in \texttt{MSSM.cpp} or to use what is already there to read processes from text files.
More detail about this can be found in the documentation of the function \texttt{computeAndAddToLibFromList}, described afterwards. 
It should be noted that it is not possible, at least in the current version of \MARTY, to add a subset of processes in a previously generated library.
So, if the user wants to add new quantities to an existing library, a new generation of the library is required.

For what concerns the usage of the numerical library, the raw output of \MARTY is not meant to be particularly flexible. 
Only recently, an index of its functions is produced automatically, hence available automatically in the self-generated numerical library.
For our purposes, this is not enough, so the \DarkPack package provides a super-set of \MARTY's numerical library, which allows for a more 
intuitive usage and additional features, such as the running of SM parameters.
The main features are described below. 

Notably, we have been able to enhance the flexibility, automatically generating a hash table (i.e.~a \lstinline{std::unordered_map}) whose keys are strings unique for each 2 to 2 process and whose values are a \lstinline{std::tuple} with all the relevant quantities for computing the sum of the squared amplitudes for it. 
Therefore, only for the processes present in the hash table it is possible to compute the relevant quantities: we describe how to do it in the example files.
We would like to point out that this procedure can be used to put in the libraries also other observables, even unrelated to the 2 to 2 processes,
such as specific decay widths or Wilson coefficients. Within this framework, more potentialities of \MARTY will be exploited in the future.

As far as input values are concerned, in this release we provide algorithms to read from \texttt{.lha} \cite{Skands:2003cj,Allanach:2008qq} files.
Decay widths at the tree level and mass spectra can be computed using \MARTY if needed, otherwise, the necessary quantities have to be provided among the input parameters.

In the \lstinline{auxiliary_library/script_mssm2to2} directory some example files are provided to show how to use the main features of the code in practice. The users are invited to contact the authors if they need some more examples or features.
We provide:
\begin{itemize}
 \item \lstinline{example_1_single_process.cpp}\\
       An example file that shows how to initialise a single process and do some calculations.
 \item \lstinline{example_2_running.cpp}\\
       An example file that shows how to easily run the Standard Model parameters.
 \item \lstinline{example_3_process_vector.cpp}\\
       An example file that shows how to efficiently deal with vectors of processes, by computing an inclusive cross-section.\\
 \item \lstinline{example_4_dWeff.cpp}\\
       An example file that shows how to compute $\displaystyle\dv{\Weff}{\cos(\theta)}$\footnote{See \cite{Gondolo:1998ah} for its definition.}.
 \item \lstinline{example_5_relic.cpp}\\
       An example file that shows how to compute the relic density given different QCD equation-of-state models.
\end{itemize}

\section{Main features of the \MARTY code}\label{sec:marty}

For the sake of simplicity and clarity, we defined the following aliases:
\begin{lstlisting}[language=C++]
    using Process = std::vector <mty::Insertion>;
    using Processes = std::vector <Process>;
\end{lstlisting}
In fact, a vector of \lstinline{Insertion} contains the information for a specific process. 
We furthermore define the extension \texttt{.psiso}, for a text file containing the names of 2 to 2 processes in the \SuperIso convention, as well as the new type
\begin{lstlisting}[language=C++]
    typedef struct
    {
        Process process;
        mty::Order order; // Options mty::Order::TreeLevel, 
                             //           mty::Order::OneLoop 
        bool leading_order;
        mty::gauge::Type Wgauge;   
    }Process2to2ToCompute;
\end{lstlisting}
that contains a process, information about the order to which its amplitude will be computed, and the choice of the gauge for the $W$ boson\footnote{
This needs to be done because a very small number of squared amplitudes can be numerically computed to be negative in the Feynman gauge. Computing them in the unitary gauge is a way to fix this inconsistency.}.

A list of the main functions that we implemented in \MARTY is given below:
\begin{enumerate}
 \item  \lstinline{std::string processName(Process const &proc)} \\
       This function takes as input a specific process, and returns a string that corresponds to a name for this process.
       This function can serve many purposes. In \DarkPack we use it to assign a name to the functions in the numerical library.
 \item \begin{lstlisting}[language=C++]
std::string generateCorrespondance(
  std::vector<Particle> psm, 
  std::vector<Particle> pbsm, 
  const std::string filename="smBsm.h") 
       \end{lstlisting} 
       This function takes as first argument a vector of \texttt{Particle} 
       that corresponds to the list of the particles
       to be considered as SM, as second argument a vector of \texttt{Particle} 
       that corresponds to the list of the
       particles to be considered as BSM, and as third optional argument a string \texttt{filename}.
       This function generates the file \texttt{auxiliary\_library/<libname>/filename} with the structure of a \texttt{C++} header.
       It will contain the list of the SM and the BSM particles that the numerical library needs to know.
 \item \begin{lstlisting}[language=C++]
int computeAndAddToLibFromList( mty::Model  &model, 
  mty::Library &lib,                                
  std::vector<Process2to2ToCompute> listofprocs,    
  std::string nameSmBsmFile="smBsm.h"               
  )
       \end{lstlisting}
       This function takes as input:
       \begin{itemize}
        \item a model,
        \item a library for the output,
        \item the list of processes to be computed and added in the library,
        \item the name of the file created with \texttt{generateCorrespondance}.
       \end{itemize}
       This function computes the amplitude of each \lstinline{Process2to2ToCompute} in the input vector within the specified gauge and to the specified order.
       If the boolean \lstinline{leading_order} is set to \lstinline{true},
       the information in \lstinline{order} is neglected, and the calculation
       stops at tree-level if the amplitude is non-zero, otherwise 
       it is performed at one-loop.
       Then such an amplitude is squared and saved in the library \lstinline{lib}. 
       This function also provides the creation of some auxiliary files in the numerical library, such as the content of \lstinline{correspondance.h},
       described in what follows.

 \item \begin{lstlisting}[language=C++]
int addFromFile(
  mty::Model &model, 
  Processes &processes,
  std::vector<std::string> &names,
  std::vector<std::string> &namesSuperiso, 
  std::string filename)
       \end{lstlisting}     
       This function has been written specifically for the MSSM. The inputs are:
       \begin{itemize}
        \item the model (MSSM),
        \item a vector of processes,
        \item two vectors of strings, that will contain the names of the processes in two conventions respectively: the one defined in \texttt{processName}, and the one used in \SuperIso,
        \item a string, which is the input file name.
       \end{itemize}
       This function reads the file \lstinline{filename}, which is a \texttt{.psiso} file. For each process name, this function reconstructs the particle content and puts in \lstinline{processes} the process, in \lstinline{names} their names in the \lstinline{processName} convention, and in 
       \lstinline{namesSuperiso} their names in the \SuperIso convention.
       The function returns the number of read processes in case of success.\footnote{For testing purposes, we added to this function the generation of a file \lstinline{processes_chep.txt}. This file has two columns and a row for each process.
       The first element of the row is the process name in the \texttt{CalcHEP} convention \cite{BELYAEV20131729} and the second row is the corresponding name in the \SuperIso convention.}
\end{enumerate}

The other functions are documented in the comments of the code.

\section{Main features for the numerical library}\label{sec:numerical}
We list below the auxiliary functions in the numerical library. In what follows, we call this library \lstinline{bsm2to2}. 
Functions specific to the MSSM described here will have explicitly the namespace \lstinline{mssm2to2}.
In order to avoid possible conflicts, most of the functions in the libraries are defined under the namespace of the name of the library.

\subsection{Global variables and generic functions}
Let us start by describing the main content of the header \lstinline{correspondance.h}, included in the other headers listed in the following.
Here are defined an enumeration, global variables and functions, in the namespace \lstinline{bsm2to2::corr}. 
The most relevant contents of this header are:
\begin{enumerate}
 \item \lstinline{enum Part_t} \\
       Such an enumeration defines particles and starts from 1.
 \item \lstinline{int SIZEPHYSICALSM}\\
       This variable contains the number of particles to be considered in the Standard Model (SM).
 \item \lstinline{int SIZEPHYSICALBSM}\\
       This variable contains the number of particles to be considered as Beyond the Standard Model (BSM).
 \item \lstinline{int TOTAL_PARTICLES}
       This variable contains the total number of particles in the model.
 \item \lstinline{std::array<int,SIZEPHYSICALBSM> bsm_particles}\\
       This array contains all the integer values of all the BSM particles. It is possible to use it
       to cycle through BSM particles.
 \item \lstinline{std::array<int,SIZEPHYSICALBSM> sm_particles}\\
       This array contains all the integer values of all the SM particles. It is possible to use it
       to cycle through SM particles.
 \item \lstinline{std::array<std::string, TOTAL_PARTICLES+1> part_names} \\
       This array is defined to have the name of each particle, corresponding to the enumeration.
 \item \lstinline{std::array<bool, TOTAL_PARTICLES+1> isboson} \\
       An element is \lstinline{true} if the corresponding particle is a boson.
 \item \lstinline{std::array<double,TOTAL_PARTICLES+1> part_charge}\\
       The elements of this array are the electric charges of each particle.
 \item \lstinline{std::array<double,TOTAL_PARTICLES+1> part_dof}\\
       The elements of this array are the degrees of freedom of each particle.
\end{enumerate}
In \lstinline{correspondance.h} is included also the header \lstinline{params_new.h}, where the type \lstinline{struct bsm2to2::Param_t} is defined. It inherits the members of the type \lstinline{struct bsm2to2::param_t}, different for each library because automatically generated by \MARTY, and it adds some quantities useful for the running. This is the type of variables that we use to handle the input parameters.

\subsection{Input reading and manipulation}
Since each model has its own parameters, the input management has to be handled by the users accordingly to their needs. In this paragraph, we describe the functions which we defined in order to work within the MSSM.

The headers we use are: 
\begin{itemize}
 \item \lstinline{leshouchesfromsuperiso.h}:  its functions are defined under the namespace \lstinline{mssm2to2::superisosupport},
 and they allow us to read from a \texttt{.lha} file calling \SuperIso's routines;
 \item \lstinline{leshouchesfrommarty.h}: its functions are defined under the namespace \lstinline{bsm2to2::readmodule}, and they allow us to read from a \texttt{.lha} file calling the \texttt{lha} extension native in the \MARTY installation. 
\end{itemize}
For a general use, the library to refer to in order to modify the input management is \lstinline{leshouchesfrommarty.h}. The library \lstinline{leshouchesfromsuperiso.h} is useful while dealing with the MSSM only.

We list here the main functions:
\begin{enumerate}
 \item \begin{lstlisting}[language=C++]
int superisosupport::InitInterfaceStruct(
    const parameters *const param, 
    param_t &input)      
       \end{lstlisting}
       This function takes as inputs a pointer to a variable of type \lstinline{parameters}, that contains the \SuperIso inputs, and an address to a \lstinline{param_t}. It copies the values contained in the \SuperIso inputs to the structure \lstinline{param_t}. In practice, the \SuperIso structure is read using the \SuperIso routines\footnote{They can be found in \lstinline{leshouches.c} and \lstinline{leshouches.h}.} 
       and then this function is called to handle the inputs in the numerical library.
 \item \begin{lstlisting}[language=C++]
int superisosupport::InitInterfaceStruct_Full(
    const parameters *const param, 
    Param_t &input)      
       \end{lstlisting}
       This function calls \lstinline{InitInterfaceStruct} and initialises the other members of the \lstinline{Param_t} variable given as input.
 \item \begin{lstlisting}[language=C++]
struct parameters superisosupport::ReadLHA(
    Param_t &input, 
    const char * name, 
    int *err)
       \end{lstlisting}
       This function uses the previous one to read the \texttt{.lha} file with the path \lstinline{filename}, and saves the data in \lstinline{input}.
       It returns a variable of type \lstinline{struct parameters}, which is the corresponding structure of the data used in \SuperIsoRF.
       If there is no error in the process, \lstinline{err} is set to zero.
 \item \begin{lstlisting}[language=C++]
Param_t readmodule::ReadLHA(
    const std::string filename);
       \end{lstlisting}
       This function reads the \texttt{.lha} file with the path \lstinline{filename} and returns a \lstinline{bsm2to2::Param_t} whose members are filled with data provided in the input file. 
       If a required SM input is not provided, it is filled with the default Particle Data Group  (PDG) \cite{Workman:2022ynf} value. 
       In the case of the MSSM, we have chosen the PDG of 2016 to ensure compatibility with \SuperIsoRF.
       This choice can be easily overridden if the user wants to use other values.\footnote{
       By the appropriate replacement of the line \lstinline{using namespace mssm2to2::pdg2016Value;} in 
       \lstinline{auxiliary_library/macros.h}, following the namespaces defined in 
       \lstinline{auxiliary_library/params_new.h}.
       }
\end{enumerate}

The other functions are documented in the comments of the headers.

Note that no further manipulation is done on the values, except for the running of the Standard Model parameters $m_t$, $m_b$, $\alphastrong$. 
This means that, if no specific code is written, the input file has to provide quantities such as the mass spectrum computed externally. 

It is in fact possible to compute the mass spectrum inside the program.
In this regard, we invite the interested users to read the \MARTY's documentation to have the full details.\footnote{See for example
\url{https://marty.in2p3.fr/doc/marty-manual.pdf}.}
In \ref{app:spectrum} we give the most relevant information on spectrum calculation.

\subsection{Handling the running of SM parameters}
Below we list the functions for the running of SM quantities, i.e.~the strong coupling constant $\gstrong$, and the top and the beauty quark masses. 
This is handled via a structure called \lstinline{RunningSM}, accessible after including \lstinline{RunningSM.h}. 
We used the code provided in \SuperIsoRF to define the methods of this class. 
Here we list its main public methods and elements:
\begin{enumerate}
 \item \lstinline{RunningSM(const Param_t &input)}\\
       This constructor builds the class variable starting from the values stored in a \lstinline{Param_t} variable.
 \item \lstinline{RunningSM(void)}\\
       This constructor builds the class variable from the default PDG values.
 \item \begin{lstlisting}[language=C++]
enum ParticlesList : short int {UP, DOWN, STRANGE, 
    CHARM, BEAUTY, TOP, EL, MU, TAU, NUE, NUMU, NUTAU, 
    GLUON, W, Z, PHOTON, HIGGS}   
       \end{lstlisting}
       It is an enumeration that contains all the physical particles in the Standard Model.
 \item \lstinline{double GetMbPole()}\\
       Returns the b-quark pole mass.
 \item \lstinline{double GetTopPoleMass()}\\
       Returns the t-quark pole mass.
 \item \lstinline{double AlphaStrong(double Q, double mtop)}\\
       Returns the value of $\alphastrong$ at the energy \lstinline{Q}, taking \lstinline{mtop}  as the value for the top pole mass.
 \item \lstinline{double GetMcPole()}\\
       Returns the charm pole mass computed from $m_c(m_c)$.
 \item \lstinline{double GetMbMb()}\\
       Returns $m_b(m_b)$.
 \item \lstinline{double GetMtopMtop()}\\
       Returns $m_t(m_t)$.
 \item \lstinline{double GetQuarkMass(enum ParticlesList part, double Qf)}\\
       Takes a quark as enumeration and an energy scale \lstinline{Qf}. 
       The return value is the mass of the particle at the scale \lstinline{Qf}.
 \item \begin{lstlisting}[language=C++]
void HandleParamRunning(
    Param_t &input, 
    const double Q)
        \end{lstlisting}
        This method performs the running of the parameters at the scale \lstinline{Q} and saves the results in \lstinline{input}. 
 \item \lstinline{void RunLightQuarks(bool x=true)}\\
       This method changes the default behaviour of \lstinline{HandleParamRunning}.
       After its calling, if {\tt x} is {\tt true}, the running of the down, up and strange masses is enabled. Otherwise,
       the running of the down, up and strange masses is disabled.
 \item \lstinline{void RunCharmMass(bool x=true)}\\
       This method changes the default behaviour of \lstinline{HandleParamRunning}. 
       After its calling, if {\tt x} is {\tt true}, the running of the charm mass is enabled. Otherwise,
       the running of the charm mass is disabled.
 \end{enumerate}

\subsection{Functions related to 2 to 2 processes}
We describe here the functions and the methods dedicated to the computation of 2 to 2 squared amplitudes, cross sections, and contributions to $W\ped{ eff}$ (see \cite{Gondolo:1998ah} for its definition).
These functions and methods are accessible by including
\lstinline{process.h}. 
In this header, the class \lstinline{Process2to2} is defined. 
In order to understand how the public methods work, we remark that the main members of this class are the private ones
\begin{lstlisting}[language=C++]
 csl::InitSanitizer<int> p[4];
 csl::InitSanitizer<bool> ap[4];
\end{lstlisting}
They are two arrays of size 4, because the class is intended for a process of the kind $1, 2 \to 3, 4$.
\lstinline{p[i]} contains the \lstinline{i}-th field as enumeration, while the \lstinline{ap[i]} is \lstinline{true} or \lstinline{false}, depending on whether the \lstinline{i}-th entry is a particle or its antiparticle.
This is always specified, even for a particle which is its own antiparticle.
It can be important to know that, when a process is filled, the order of the particles may change. 
In fact, after the finalisation of a process, an algorithm determines whether the sum of the squared amplitudes of the process is in the library.
If it is in, the particles are re-ordered accordingly to the order they appear in the function that computes the sum of the squared amplitudes.
This is relevant to verify if some quantities related to kinematic parameters have to be computed: in such a case, it is necessary to 
calculate them accordingly to the order in which they appear in the \lstinline{Process2to2} variable.
Furthermore, the following type is defined:
\begin{lstlisting}[language=C++]
 struct Insertion
 {
   int field;
   bool part;
  
   Insertion(int i, bool b=true)
   {
     field = i;
     part = b;
   };
 };
\end{lstlisting}
in order to specify the particles that enter a process. In practice, it is possible to construct a variable of this type by assigning an integer, that corresponds to the field, or by assigning a two element list, with an integer and a boolean, where the boolean determines if it is a particle or an antiparticle.
The main public methods are the following ones:
\begin{enumerate}
 \item \lstinline{Process2to2()}\\
       This is a constructor: it creates an empty process.
  \item \begin{lstlisting}[language=C++]
Process2to2(const std::array<Insertion,4> &v);
       \end{lstlisting}
       This constructor constructs the class corresponding to the process with incoming (outgoing) particles 
       whose field is in the first (last) two elements of \lstinline{v}.
 \item \begin{lstlisting}[language=C++]
Process2to2(const std::array<int,4> &p,
              const std::array<bool,4> &ap)
       \end{lstlisting}
       This constructor takes as input two arrays of size 4. 
       It constructs the class corresponding to the process with incoming (outgoing) particles whose field is in the first (last) two elements of \lstinline{p}.
       The second argument corresponds to \lstinline{true} (\lstinline{false}) if the element is a particle (antiparticle).
 \item \begin{lstlisting}[language=C++]
short int set(
    short int n, 
    const int &ip, 
    const bool iap)
 \end{lstlisting}
       This function sets the \lstinline{n}-th particle with field \lstinline{ip} and (anti)particle \lstinline{iap}.
 \item \lstinline{std::string getName()}\\
       Returns the name of the process.
 \item \lstinline{std::string getMname()}\\
       Returns the name of the process according to the convention in \MARTY.
 \item \lstinline{std::string getSname()}\\
       Returns the name of the process according to the convention in \SuperIsoRF.
 \item \lstinline{double getMass(short int i)}\\
       Returns the mass of the \lstinline{i}-th particle in the process.
 \item \begin{lstlisting}[language=C++]
double getSumSquaredAmpl(
    Param_t &input, 
    const double &sqrts, 
    const double &ctheta)
       \end{lstlisting}
       Returns the sum of the squared amplitudes for the process, with the numerical inputs contained in \lstinline{input}, the centre of mass energy \lstinline{sqrts}, and the cosine of the angle between particle 1 and particle 3 \lstinline{ctheta}.
 \item \begin{lstlisting}[language=C++]
double getAvgSquaredAmpl(
    Param_t &input, 
    const double &sqrts,    
    const double &ctheta)
       \end{lstlisting}
       Returns the average of the squared amplitudes for the process, with the numerical inputs contained in \lstinline{input}, the centre of mass energy \lstinline{sqrts}, and the cosine of the angle between particle 1 and particle 3 \lstinline{ctheta}.
 \item \begin{lstlisting}[language=C++]
double getDiffWeffContrib(
    Param_t &input, 
    const double &sqrts, 
    const double &ctheta)
       \end{lstlisting}
       Returns the contribution to $\dv{W\ped{eff}}{\cos(\theta)}$ for the process, defined as (see e.g.~\cite{Arbey:2009gu, Gondolo:1998ah}):
       \begin{equation}\label{eqn:dweffdcos}
        \qty(\dv{W\ped{ eff}}{\cos(\theta)})_{1,2\to3,4} = \frac{f\ped{ CP} p_{12}p_{34}}{S_{f34}\sqrt{s}\peff}\sum_{\text{all d.o.f.}} \abs{M}^2
       \end{equation}
       where 
       \begin{enumerate}
         \item $f\ped{ CP} = 2$ if $\bar 1 \bar 2 \to \bar 3 \bar 4$ is allowed, 
               otherwise it is 1
         \item $S_{f34} = 2$ if 3 and 4 are the same field and the same particle 
               or anti-particle, otherwise it is 1
         \item \begin{equation*}
                p_{ij} = \frac{\sqrt{\qty(s- (m_i + m_j)^2)\qty(s - (m_i -m_j)^2)}}{2\sqrt s}
               \end{equation*}
         \item \begin{equation*}
                \peff = \frac{1}{2}\sqrt{s -4m_{\rm LBSM}^2}
               \end{equation*}
            with $m_{\rm LBSM}$ the mass of the lightest stable new particle.
       \end{enumerate}
       For the computation the numerical inputs
       contained in \lstinline{input} are used, together with the centre of mass energy \lstinline{sqrts}, and the cosine of the angle between particle 1 and particle 3 \lstinline{ctheta}.
 \item \begin{lstlisting}[language=C++]
double getDiffCrossSection(
    Param_t &input, 
    const double &sqrts, 
    const double &ctheta)
       \end{lstlisting}
       Returns the differential cross section for the process, with the numerical inputs contained in \lstinline{input}, the centre of mass energy \lstinline{sqrts}, and the cosine of the angle between particle 1 and particle 3 \lstinline{ctheta}.
 \item \begin{lstlisting}[language=C++]
double getTotalCrossSection(
    Param_t &input, 
    const double &sqrts,
    double *discr=nullptr);
       \end{lstlisting}
       Returns the total cross section for the process, with the numerical inputs contained in \lstinline{input}, and the centre of mass energy \lstinline{sqrts}.
       The optional argument, if specified, allocates in \lstinline{discr} the relative error of the integral. 
       For details, see section \ref{sec:integrals}.
\end{enumerate}
Other methods of this class are defined in \ref{app:runmilti}.

\subsection{Computation of $\expval{\sigma v}$}
Whereas it is possible to use what has been described so far to link \DarkPack to any external tool to compute $\expval{\sigma v}$,
we provide a class named \lstinline{AvgSvCalculator} for its computation, 
defined in \lstinline{avgsvcalculator.h}, that allows for the computation of the average annihilation rate $\Weff$ defined 
in \cite{Gondolo:1998ah} and also $\expval{\sigma v}$.
More generally, its purpose is the computation of quantities for a list of processes in a more efficient way in terms of performance.
In order to understand the way it works, it would be useful to know that it has the following private members:
\begin{lstlisting}[language=C++]
 RunningSM run;
 Param_t input;
 std::shared_ptr<std::vector<Process2to2>> p_ptr;
\end{lstlisting}
The initialisation of \lstinline{std::vector<Process2to2>} takes time: this is the reason we use a shared pointer and defined a
copy constructor for this class.
Its public methods are the following:
\begin{enumerate}
 \item \lstinline{AvgSvCalculator(const Param_t &param)}\\
       It is the default constructor. It copies the content of \lstinline{param} into \lstinline{input}, then it constructs \lstinline{run}, and it allocates the vector \lstinline{p} with all the possible 2 to 2 
       processes of the kind BSM + BSM to SM + SM, whose sum of squared amplitudes is available in the library, avoiding duplicates.
       Since initialising a single process needs more passages, the creation of a variable with this constructor takes time. For this, we recommend to create a variable as global and then copy its content in functions if necessary.
 \item \lstinline{void runAtScale(const double Q)}\\
       This method does all it takes to have all the quantities run at the energy scale \lstinline{Q}.
 \item \lstinline{void setWeffcuts(bool x);}\\
       If \lstinline{x} is \lstinline{true}, the cuts on $\Weff$ are enabled, otherwise they are disabled.
       By default they are enabled after the call of the constructor, so you must call this method to disable them
       after the constructor call.
 \item \lstinline{double getWeff(const double sqrtS)}\\
       This method returns $g^2\ped{ LNP}W\ped{ eff}$, with $g\ped{ LNP}$ the number of degrees of freedom of the lightest stable NP particle,
       at the centre of mass energy \lstinline{sqrtS}. 
       In this method, the lightest stable NP particle is determined from the content of the given \lstinline{input}.
 \item \lstinline{void tabulateValues(const double &sqrtSmax, const size_t &Nmax)}\\
       Creates a table of values for $\Weff$, with maximum energy \lstinline{sqrtSmax} and size of reference \lstinline{Nmax}.
       Such a table is a private member of the class.
 \item \lstinline{double getAverageSigmav(const double &T)}\\
       Calculates the value of $\expval{\sigma v}$ in $\unit{GeV^{-2}}$ at the temperature $T$. If necessary, this method allocates the table of values for $\Weff$
       with default parameters.
\end{enumerate}

\subsection{Relic density calculation}

We also provide the tools to compute the relic density for models with a single dark matter candidate. 
Our algorithm is the same as the one in \SuperIsoR.
Hence, a structure \lstinline{Relicparam_t} is defined in the header \lstinline{relicparam.h}, that works as the structure \lstinline{relicparam}
in \SuperIsoR. For the sake of simplicity we translated the functions in \SuperIsoRF into methods.
For their documentation, we refer the user to \cite{Arbey:2018msw}.
Finally, a class \lstinline{BoltzmannSolver} is defined, child of \lstinline{AvgSvCalculator} and \lstinline{Relicparam_t}.
As for \lstinline{Relicparam_t}, most of its methods have the same name of functions in \SuperIsoR.
Fur such methods, the documentation is found in \cite{Arbey:2018msw}.
We describe in what follows the constructors and the main methods:
\begin{enumerate}
 \item \lstinline{BoltzmannSolver(const Param_t &param, int qcdmodel=2)}\\
       This constructor instantiates an object by calling the parent constructor \lstinline{AvgSvCalculator(const Param_t &)}
       that builds the parent class, and chooses the QCD model 2 by default.\footnote{for the definition of the QCD models
       we refer to \cite{Arbey:2018msw}.}
 \item \lstinline{BoltzmannSolver(const AvgSvCalculator &sop, int qcdmodel=2)}\\
       This constructor instantiates an object by calling the copy constructor for the class \lstinline{AvgSvCalculator}
       that builds the parent class, and chooses the QCD equation-of-state model 2 by default.
 \item \lstinline{double relic_density()}\\
       This method computes the relic density with the default \SuperIsoRF algorithm.
 \item \lstinline{void print()}\\
       This function prints the list of processes, the table for $\Weff$ and the values of the members inherited by \lstinline{Relicparam_t}.
       
\end{enumerate}

\section{Comparison with other softwares}\label{sec:integrals}

We validated our code within the MSSM, testing 3430 processes of the type SUSY + SUSY $\to$ SM + SM. These processes are the ones that can be directly found in the \texttt{mssm2to2} library in the repository. 
We compared our results for the sums of the squared amplitudes and the contribution to $\dv{\Weff}{\cos(\theta)}$
with the ones from \SuperIsoR \cite{Arbey:2018msw, Arbey:2011zz, Arbey:2009gu}, and our results for total and differential cross sections with the ones from \texttt{micrOMEGAs}/\CalcHEP \cite{Belanger:2018ccd, BELYAEV20131729}. 
In this way, we validated the behaviour of \MARTY's MSSM library \cite{Uhlrich:2020ltd} and also we performed numerical tests.

The comparison of the results of the total cross sections with \texttt{micrOMEGAs}/\CalcHEP has shown that there are some cross sections (around 9\permil) for which the integration algorithm produces numerically unstable results with any software.
This could be due to the presence of resonances, however, the full understanding of the underlying reason is not achieved yet. 
In order to be able to control such a behaviour without affecting too much performance, we integrate the differential cross sections with the Gauss-Legendre method at two different orders and we compute the discrepancy.
If the value is not accurate enough we pass to the next polynomial order and we compare it with the previous one. When the 9th order is reached and convergence is not reached yet, the trapezoidal rule is used, since it is a method whose error can be controlled and whose increment in order is done without losing the previous evaluations. 
Therefore, subsequent evaluations with the trapezoidal rule are made, and if convergence is reached the algorithm stops. 
If the number of intervals exceeds 256, the last value with the last error is taken as the result.\footnote{The uncertainty on the numerical integration is provided by passing a \lstinline{double*} to the integration function.}
We analysed the behaviour of this method, concluding that the trapezoidal rule made converge 16\% of the cross sections that do not converge with the Gauss-Legendre method, achieving the 
9\permil\ of non-convergence mentioned earlier.

\begin{figure}
    \centering
    \includegraphics{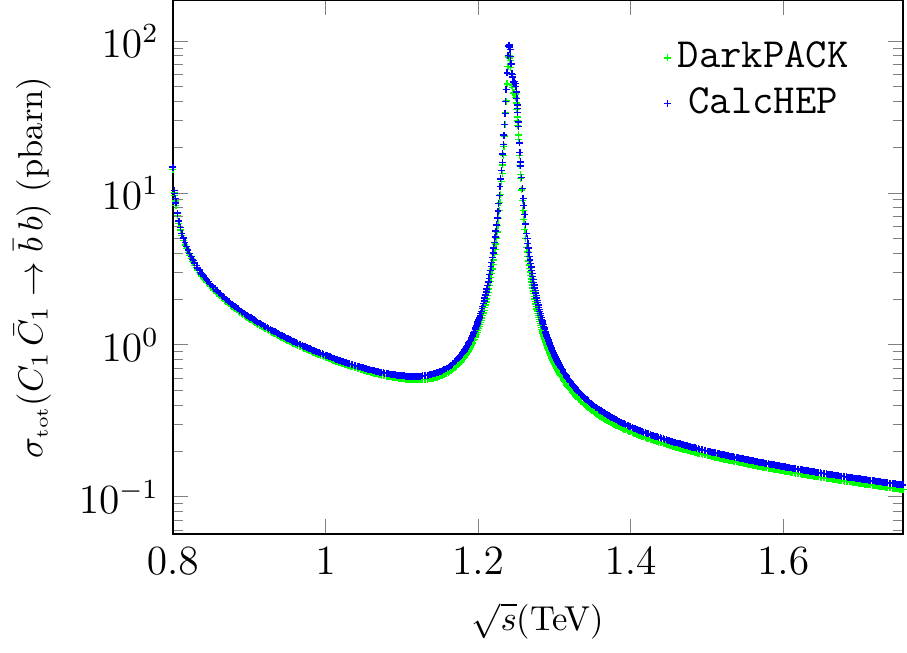}
    \caption{Total cross section for $C_1 \,  \bar C_1 \to \bar b \, b $ obtained with \DarkPack and \CalcHEP. The two are in agreement.}
    \label{fig:c1c1barbbarb}
\end{figure}

\begin{figure}
    \centering
    \includegraphics{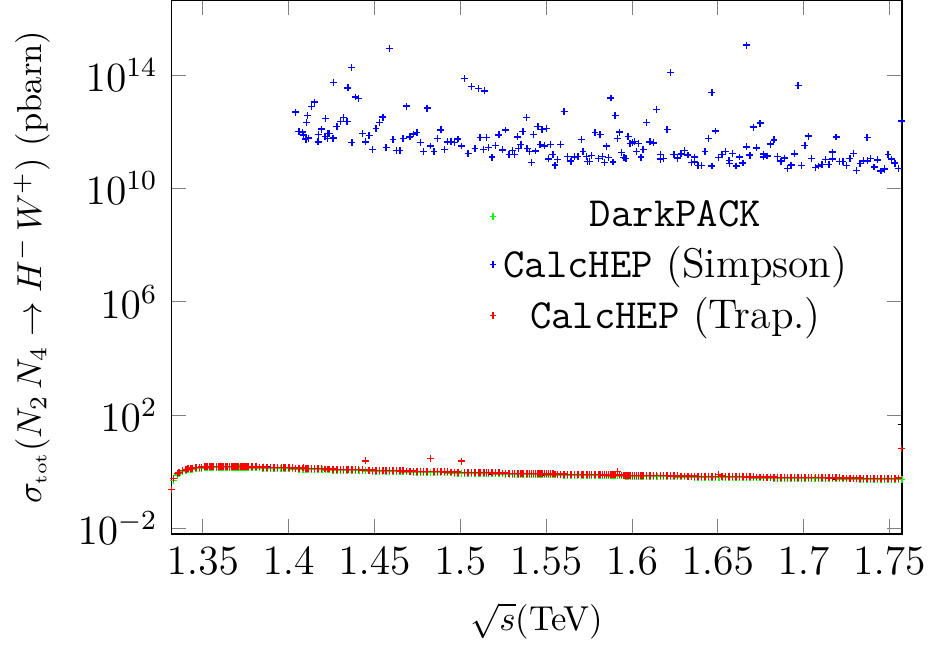}
    \caption{Total cross section for the process $N_2 \,  N_4 \to H^- \,  W^+$ computed with \DarkPack, with \CalcHEP's default integration and with a trapezoidal rule integration in \CalcHEP.
    Note that the Simpson's rule fails to several orders of magnitude.}
    \label{fig:o2o4hcbarw_plot}
\end{figure}

\begin{figure}
    \centering
    \includegraphics{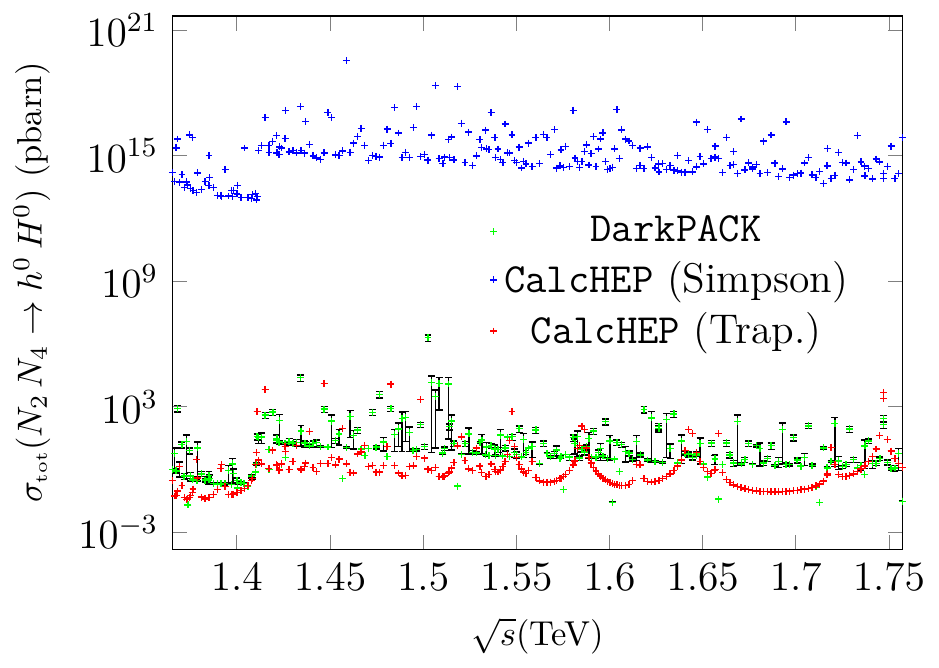}
    \caption{Total cross section for the process $N_2 \, N_4 \to h^0\,  H^0$ computed with \DarkPack, with \CalcHEP's default integration and finally with the trapezoidal rule integration in \CalcHEP.
    Note that Simpson's rule fails to several orders of magnitude, and in \DarkPack the trapezoidal rule on 256 intervals does not converge.}
    \label{fig:o2o4hhh_plot}
\end{figure}

\begin{figure}
    \centering
    
    \includegraphics[width=0.33\textwidth]{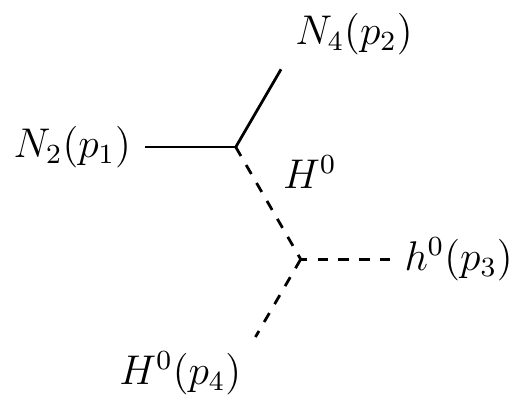}\includegraphics[width=0.33\textwidth]{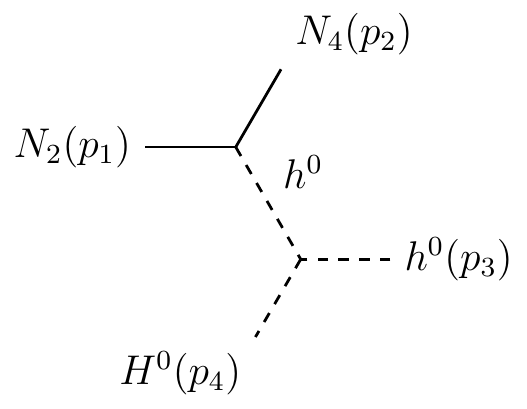}\includegraphics[width=0.33\textwidth]{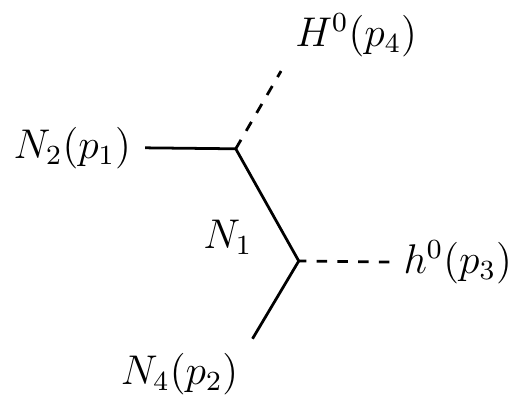}
    
    \includegraphics[width=0.33\textwidth]{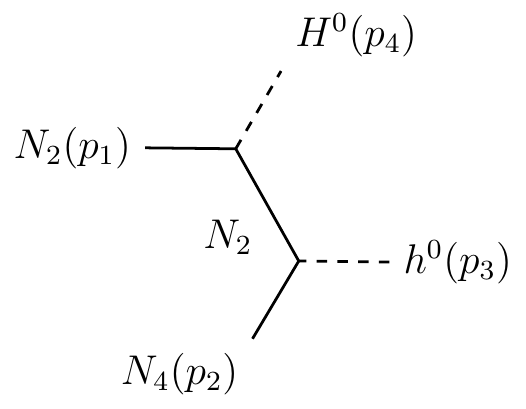}\includegraphics[width=0.33\textwidth]{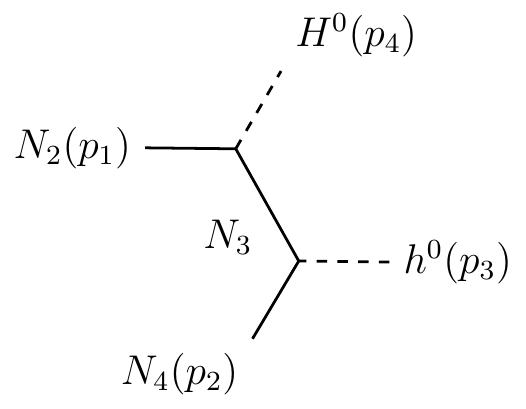}\includegraphics[width=0.33\textwidth]{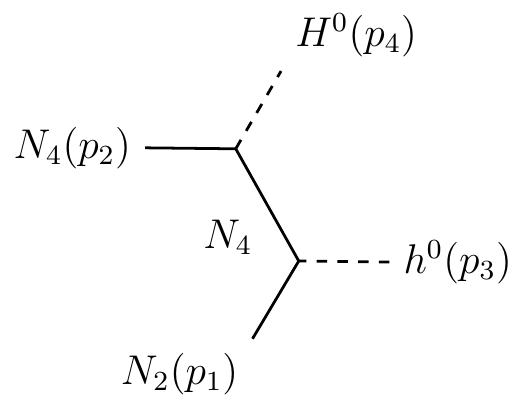}
    
    \caption{Feynman diagrams contributing to  $N_2\,  N_4 \to h^0\,  H^0$ at tree-level.}
    \label{fig:o2o4hhh_feyn}
\end{figure}

\begin{figure}
    \centering
    \includegraphics{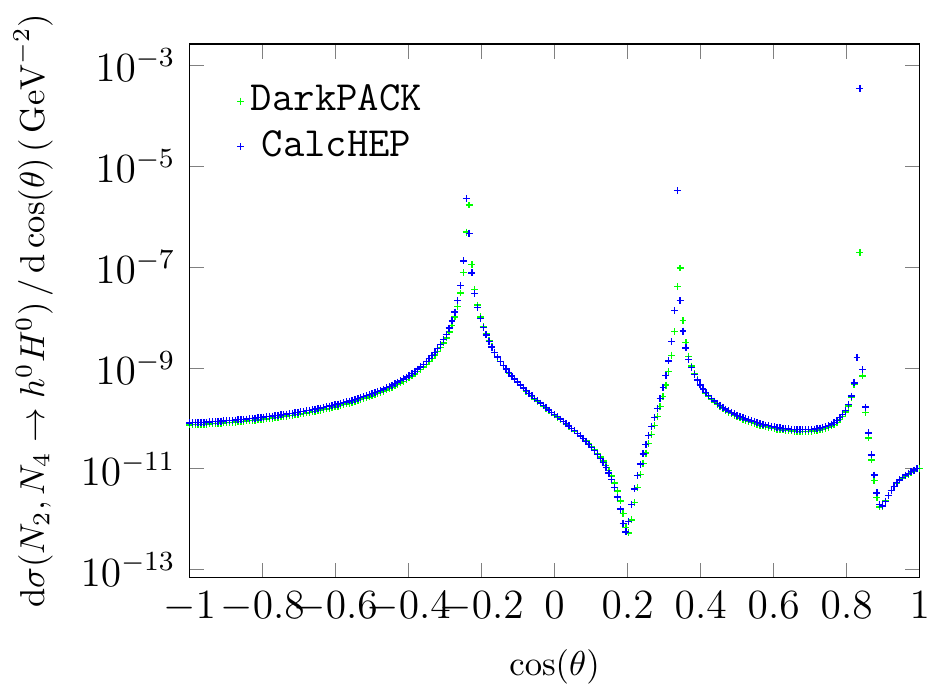}
    \caption{Differential cross section for the process $N_2 \, N_4 \to h^0 \, H^0$ at $\sqrt{s} = 1.497 \TeV$ computed with \DarkPack, and with \CalcHEP.
    Note that the two codes are in good agreement.}
    \label{fig:o2o4hhh_diff}
\end{figure}

\begin{figure}
 \centering
 \includegraphics{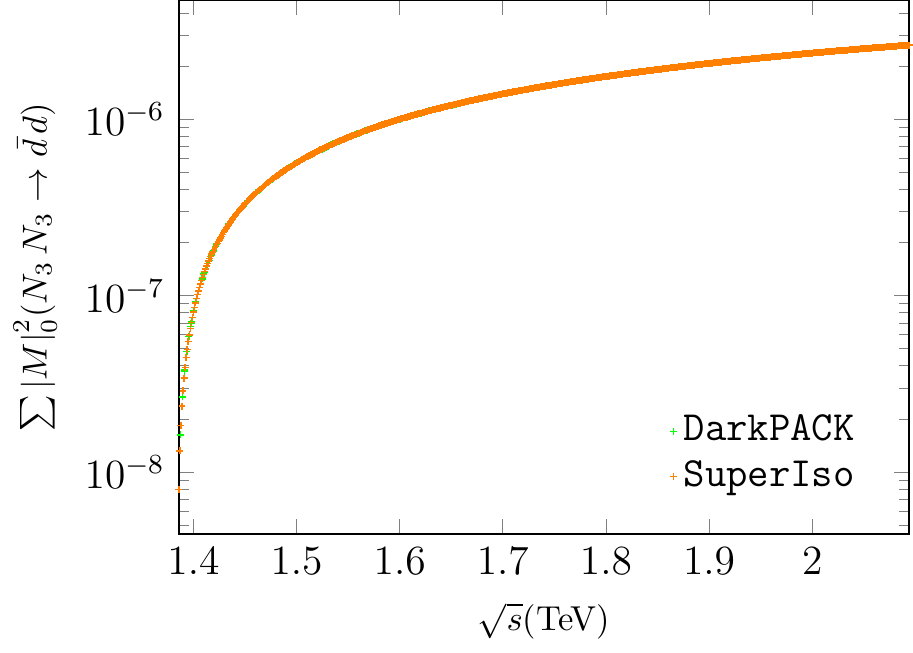}
 \caption{Sum of the squared amplitudes for the process $N_3 \, N_3 \to \bar d\,  d$ at $\cos(\theta) = 0$. \DarkPack and \texttt{SuperIso} are in agreement.}
 \label{fig:o3o3dbardM2}
\end{figure}

\begin{figure}
 \centering
 \includegraphics{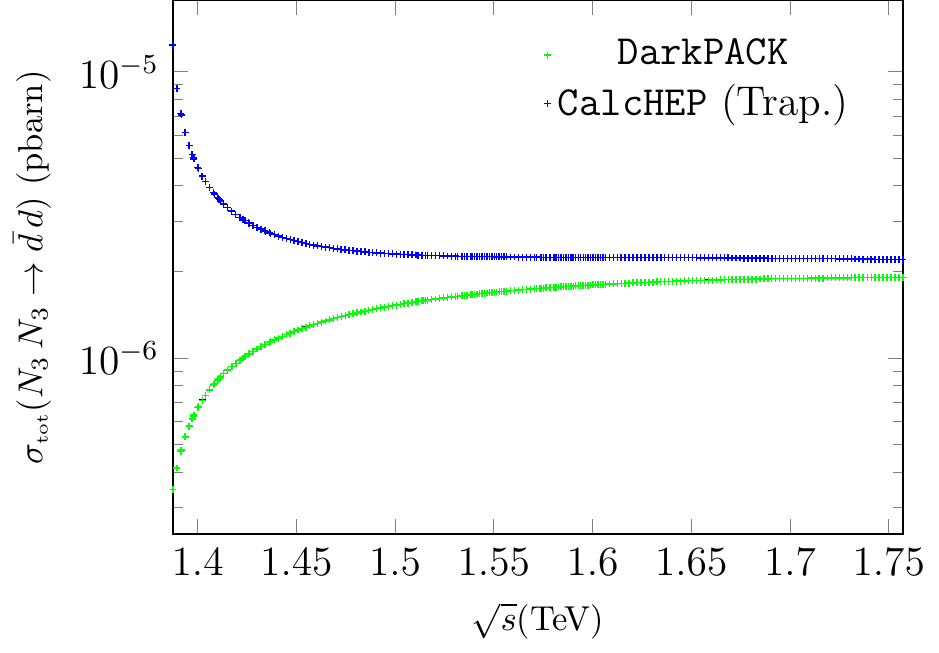}
 \label{fig:o3o3dbardxsec}
\end{figure}

The comparison with other softwares has been made by taking care of using the inputs in the same way.
In particular, we have set the CKM matrix to the identity matrix, in order to compare the results with the \SuperIsoR/\texttt{FormCalc} code, and the mass of the muon and the electron to zero for the comparison with \texttt{micrOMEGAs}/\CalcHEP.
The outcome from the comparison with \SuperIsoR is that all the sums of the squared amplitudes are in agreement, except for the heights of some peaks.
The comparison with \texttt{micrOMEGAs}/\CalcHEP shows that most of the cross sections are in agreement. An example is provided in figure \ref{fig:c1c1barbbarb}.\footnote{
In our notation, the $C_i$ corresponds to the positively charged chargino with the $i$-th lowest mass, while
$N_i$ is the neutralino with the $i$-th lowest mass.
}
The differences between the results with \texttt{micrOMEGAs}/\CalcHEP can be classified in several categories, for which we provide some sample plots:
\begin{itemize}
    \item By default, \CalcHEP uses the Simpson rule to compute total cross-sections from the differential ones. For some processes, such a rule fails to achieve a correct result. To be able to compare \CalcHEP's results with ours, we implemented the trapezoidal rule in \CalcHEP to get the same quantities. An example of such a behaviour is shown in figure \ref{fig:o2o4hcbarw_plot};
    \item In some other cases, no algorithm gives a coherent result, because 
        \begin{itemize}
            \item as mentioned earlier, the Simpson rule in \CalcHEP fails to achieve a correct result;
            \item our algorithm, while giving a numerically reasonable result, is  affected by large uncertainties and fluctuations;
            \item the trapezoidal rule in \CalcHEP converges to a numerically reasonable result, different from ours and with a negligible uncertainty and fewer fluctuations, but it presents some resonance-like peaks that are not supposed to exist. 
        \end{itemize}
        In figure \ref{fig:o2o4hhh_plot} an example of such a behaviour is shown, and the Feynman diagrams contributing to the corresponding process are shown in figure \ref{fig:o2o4hhh_feyn}.
        Amongst all the particles in the internal legs, none of them has a mass that corresponds to the higher 
        energetic peaks shown in the plot.
        An example of a differential cross section with several peaks is given in figure \ref{fig:o2o4hhh_diff}.
    \item For some processes, our results and the ones of \CalcHEP are different at the threshold.
          However, we are in agreement with \SuperIsoR. An example of this is shown in figures \ref{fig:o3o3dbardM2} and \ref{fig:o3o3dbardxsec}.
\end{itemize}

\begin{figure}
    \centering
    \includegraphics{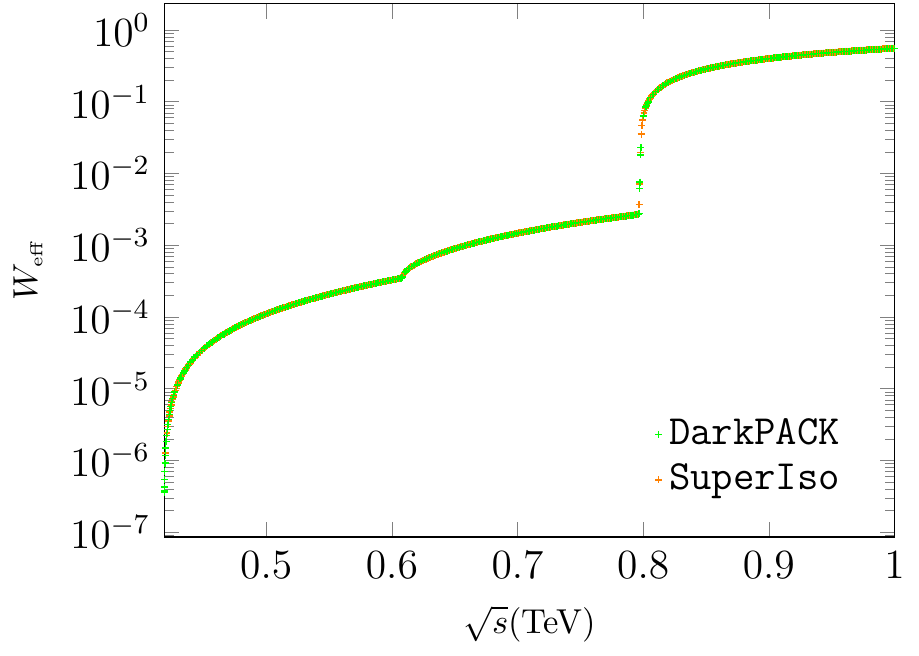}
    \caption{Comparative plot of $W\ped{eff}$ with the same data set of cross sections in the other figures.}
    \label{fig:weff}
\end{figure}

Finally, we present in figure \ref{fig:weff} $W\ped{eff}$ obtained with \DarkPack and with \SuperIsoR. 
As can be seen in the figure, the two softwares are in excellent agreement in the energy range relevant for the calculation of the relic density.
We do not show the data for other softwares, since \SuperIsoR is for example in excellent agreement with \texttt{DarkSUSY}.

The results we obtained for the computations of $\expval{\sigma v}$ and of the relic density have also been validated by a comparison
with \SuperIsoRF.

\section{Conclusions}

In this paper, we presented a new package to deal with sums of 2 to 2 squared amplitudes to the leading order in any New Physics scenario,
as well as other dark matter related quantities, such as $\expval{\sigma v}$ and the relic density.
The development of such a code addresses the needs for modularity and flexibility which can be more challenging to achieve in already existing tools, such as \FormCalc 
\cite{Hahn:2016ebn} and \texttt{CalcHEP} \cite{BELYAEV20131729}, 
which are the backends for \texttt{SuperIso Relic v4}
\cite{Arbey:2018msw} and \texttt{micrOMEGAs5.0} \cite{Belanger:2018ccd}, respectively.

Using \DarkPack, the user can more easily parallelise the calculations, since global variables have been avoided as much as possible, achieving a better time performance. 
Additional features may be added upon request of the users.
The part of this code relative to dark matter uses \SuperIsoR as a base, and adds more features and functionalities. 
In the next version of \DarkPack, already in preparation, we will provide also examples of non-supersymmetric models.
In the current release, the user has the possibility of defining custom models, within the capabilities of \MARTY, 
and to perform calculations up to the one-loop level. 

Future developments of \DarkPack will involve performance and feature improvements and upgrades. 
For instance, we plan to extend the study of the freeze-out scenario to the case of 
co-scattering (i.~e.~adding the processes of the kind BSM + SM $\to$ BSM + SM in the 
collisional term of the Boltzmann equation), and of the case in which the model 
contains more than one dark matter candidate, by generalising the Boltzmann 
equation solution algorithm to solve the system of differential equations
defined by the Boltzmann equations of each particle type taken individually.

The next versions of \SuperIsoR will benefit from this code, by
removing the hard-coded \FORTRAN code for the squared amplitudes, and consequently benefiting in modularity, with the goal of linking it with
other softwares, as well as to increase the number of predictable observables.
Furthermore, flavour physics observables can be computed and precision tests can be performed thanks to the generic version of \SuperIso \cite{Mahmoudi:2007vz,Mahmoudi:2008tp,Mahmoudi:2009zz,Neshatpour:2022fak}, providing a multi-sectorial setup to study BSM scenarios.

\section*{Acknowledgements}
We would like to thank G. Uhlrich for his support with \MARTY, and G. Bélanger and A. Pukhov for discussions about calculation within \MicrOMEGAs and \CalcHEP.

\cleardoublepage
\appendix
\section{The \SuperIso convention for names}

We describe here the \SuperIso convention for the names of 2 to 2 processes. 
This is a convention that works only for SUSY + SUSY $\to$ SM + SM type of processes.
However, we used it in \DarkPack since it is practical and since these processes are the ones we are mainly interested in. 
As described in section \ref{sec:marty}, as well as in the documentation of \MARTY, there are also other ways to define processes.

The structure of the name is of the kind \texttt{p1(bar)p2(bar)p3(bar)p4(bar)}, 
where the ``bar'' is present if the preceding field has to be taken as an antiparticle.
The list of the particles and their names can be found in Table \ref{tab:sisoconv}.
The user can check the file \texttt{processes.psiso} for the full list of the processes we tested, named in this convention.

\begin{table}[bh]\centering
 \begin{tabular}{cccc}
 \toprule
  SUSY name	&SUSY particle	&SM name	&SM particle\\
  \midrule
o1, o2, o3, o4	&the four neutralinos	&h,hh	&light,heavy Higgs\\
c1, c2	&the two charginos	&hc	&charged Higgs\\
t1, t2	&stop 1 and 2	&h3	&pseudoscalar Higgs\\
b1,b2	&sbottom 1,2	&w,z	&W,Z\\
dr,dl	&sdown right,left	&g	&gluon\\
ur,ul	&sup right,left	&a	&photon\\
cr,cl	&scharm right,left	&d,u,s,c,b,t	&quarks\\
sr,sl	&sstrange right,left	&e,m,l	&charged leptons\\
er,el	&selectron right,left	&ne,nm,nl	&neutrinos\\
mr,ml	&smuon right,left	&	&\\
l1,l2	&stau 1,2	&	&\\
ne	&electron sneutrino	&	&\\
nm	&muon sneutrino	&	&\\
nl	&tau sneutrino	&	&\\
go	&gluino	&	&\\
\bottomrule
 \end{tabular}
\caption{List of the names of the particles in the \SuperIso convention.}
\label{tab:sisoconv}
\end{table}

\section{Mass spectrum calculation}\label{app:spectrum}
Full information on mass spectra and general properties of the numerical libraries generated by \MARTY can be found in chapter 7 of its manual.\footnote{
\url{https://marty.in2p3.fr/doc/marty-manual.pdf}.} 
In this section we explain what it is done in the \texttt{MSSM.cpp} file.
The relevant part of the code, where \lstinline{lib} is the class that handles the library that will eventually be exported is this one (see listing~\ref{lst:massspectrum}).
\begin{lstlisting}[language=C++,float=tb,caption={Treatment of the mass spectrum in \MARTY.},%
captionpos=b,label=lst:massspectrum,
frame=single]
// Listing all the Physical particles
std::vector <Particle> part_0 = 
    mssm.getPhysicalParticles([&](Particle p) { 
    return p->isPhysical(); });  
std::vector <Particle> part;
for ( size_t i = 0 ; i != part_0.size() ; i++ )
{
    if(!(IsOfType<GhostBoson>(part_0[i]) || 
    IsOfType<GoldstoneBoson>(part_0[i]))) 
        part.push_back(part_0[i]);
}
// Re-fixing mass names
for ( size_t i = 0 ; i != part.size() ; i++ )
{
    if(!part[i]->getMass()->getName().empty())
    {
        part[i]->getMass()->setName("m_"+part[i]->getName());
    }
}
// In the following statement mssm is a mty::Model variable
// and lib is a mty::Library variable
lib.generateSpectrum(mssm);
\end{lstlisting}
In this way, all the physical particles are listed, and their mass terms are all grouped. The parameter that appears in the \lstinline{struct param_t} as \texttt{m\_(name of the particle)} is no longer the bare mass in this way, but the mass that has been already computed by a spectrum generator.

It is however possible to give the bare masses as starting input values in a variable  \lstinline{struct param_t params} and then call \begin{lstlisting}[language=C++]
 updateSpectrum(params);
\end{lstlisting}
if the user wants to create the whole spectrum, or just 
\begin{lstlisting}[language=C++]
 updateMassExpressions(params);
\end{lstlisting}
if the user wants only to update the mass expressions without performing the diagonalization.

\section{Running for multiple calculations at the same energy}\label{app:runmilti}

In this appendix, we want to explain how to efficiently perform the running when more quantities are computed at the same energy. 
Our choice of the whole setup for calculation has been guided by the idea of limiting as much as possible global variables, in order to give the user the possibility to easily parallelise calculation, for instance simply using the \texttt{C++} 
standard libraries.

In order to fully understand the content of this section, we will refer to what we explained in section \ref{sec:numerical}. 
A practical example about how to make use of what we describe in the following can be found in the file 
\begin{lstlisting}
 auxiliary_library/script_mssm2to2/example_3_process_vector.cpp 
\end{lstlisting}
A variable of the type \lstinline{Process2to2} has, amongst its private members, the following ones:
\begin{lstlisting}[language=C++]
 RunningSM *runptr;
 bool isRunDataExternal;
 bool isRunningExternal;
\end{lstlisting}
The default values after construction of the those members are 
\lstinline{nullptr} and \lstinline{false}.

When a function to get the sum of the squared amplitudes or the cross section is called, the default behaviour is that if \lstinline{runptr} is equal to \lstinline{nullptr}, 
a new \lstinline{RunningSM} class is allocated, the value of \lstinline{isRunDataExternal} is set to \lstinline{true}. 
This allows the destructor of the class to know it has to free the memory pointed by the \lstinline{RunningSM} member.
Furthermore, the calculation of every quantity at a given energy can be performed handling the running via \lstinline{runptr}.

However, if the user has to compute many quantities of different processes at the same energy, performing the running for each process is a waste of resources. So a class \lstinline{RunningSM} can be constructed before creating the processes, and after that one can use the methods 
\lstinline{setRunningExternal} and \lstinline{setRunningData} to perform the running once and on the defined variable. 
For instance, for the process
$N_1, N_1 \to Z, Z$:
\begin{lstlisting}[language=C++]
 RunningSM run(input);
 std::array<Insertion,4> v = {corr::N_1, corr::N_1, 
                               corr::Z, corr::Z};
 Process2to2 proc(v);
 proc.setRunningData(&run);
 proc.setRunningExternal();
 
 double Ecm=3.0e+3; // Choosing 3 TeV as center of mass energy
 run.HandleParamRunning(input, Ecm);
 double xsection = proc.getTotalCrossSection(input, Ecm);
\end{lstlisting}
where \lstinline{input} is a \lstinline{Param_t} type variable
properly initialised.
The same result will be produced by simply using
\begin{lstlisting}[language=C++]
 std::array<Insertion,4> v = {corr::N_1, corr::N_1, 
                               corr::Z, corr::Z};
 Process2to2 proc(v);
 double Ecm=3.0e+3; // Chosing 3 TeV as center of mass energy
 double xsection = proc.getTotalCrossSection(input, Ecm);
\end{lstlisting}
as shown in 
\begin{lstlisting}
 auxiliary_library/script_mssm2to2/example_1_single_process.cpp 
\end{lstlisting}
The example file we mentioned at the beginning of this appendix shows how to efficiently compute an inclusive cross section in this way.

Note that with this method the user can choose to use qualifiers in the declaration of the \lstinline{RunningSM} variable, allowing local instances to the thread and being able to coherently parallelise calculations with the \texttt{C++} standard libraries.

\cleardoublepage
\phantomsection
\bibliographystyle{unsrt}
\bibliography{database.bib}{}

\end{document}